\documentstyle{article}
\newtheorem{theorem}{Theorem}
\newtheorem{definition}[theorem]{Definition}

\newtheorem{remark}[theorem]{Remark}

\begin{document}

\title{Application of Support Vector Machine to detect an association
between a disease or trait and multiple SNP variations.}
\author{Gene Kim$\stackrel{1}{}$, MyungHo Kim$\stackrel{2}{}$\thanks{%
The order of authors are alphabetical.} \\
%EndAName
\\
19 Gage Road,$\stackrel{1}{}$ \\
East Brunswick, NJ 08816\\
\\
bernhardkim@yahoo.com$\stackrel{1}{}$\\
mkim@genecoop.com$\stackrel{2}{}$}
\date{}
\maketitle

\begin{abstract}
After the completion of human genome sequence was anounced, it is evident
that interpretation of DNA sequences is an immediate task to work on. For
understanding their signals, improvement of present sequence analysis tools
and developing new ones become necessary. Along this current trend, we
attack one of the fundamental questions, which set of SNP(single nucleotide
polymorphism) variations is related to a specific disease or trait is. For,
in the whole DNA sequence, it is known that people have different DNAs only
at SNP locations, and moreover, the total SNPs are less than 5 millions,
finding an association between SNP variations and certain disease or trait
is believed to be one of the essential steps not only for genetic researches
but for drug design and discovery. In this paper, we are going to present a
method of detecting whether there is an association between multiple SNP
variations and a trait or disease. The method exploits the Support Vector
Machine which has been attracting lots of attentions recently.
\end{abstract}

\section{An introduction}

Even though the DNA sequences are almost identical among people, they are
slightly different in the respect with appearance, height, eye color,
characters etc. Therefore it seems reasonable to believe that if there is no
environmental factor, each whole genome produce a unique person, and we
assume that the mutations at the SNP locations makes people different, in
other words, two persons with the same SNPs are identical.

The problem of determining whether a set of SNP variation cause a specific
disease or trait could be formulated as follows. For a given disease or
trait,

1. For each set of SNP variations, find its representation as a vector in a
Euclidean space.(We will discuss this in the {\bf section 3})

2. Get a systematic way of distinguishing SNP genotypes of normal people
from ones of people with the disease or trait. To do it, we will use the
Support Vector Machine introduced by Vapnik, and we will describe its idea
briefly in the next section.

We can explain why this Support Vector Machine is natural, by considering
the examples in our real life. The current law prohibits a driver who has
alcohol concentration in his/her blood over a certain number. This is an
example of representation of a degree of ''drunkenness'' by a number. When
we describe a basketball player, height and weight are mentioned most
frequently and we trust those two numbers are indices for potential as a
basketplayer. Enumeration of numbers such as (height, weight) is called a
vector in mathematics and a generalization of a number(enumeration of only
one number). Once we have a representation, we have only to find out a way
to separate into two groups. For the case of alcohol concentration, drunken
drivers or not, depending on whether the concentration is greater than the
number allowed legally. The role of the Support Vector Machine determining
the boundary of two groups.

\section{A review of Support Vector Machine}

Let $R^n$ be a $n-$dimensional Euclidean space and let $A$ and $B$ be two
sets of finite number of points in $R^n$. The basic and fundamental question
is whether there is a systematic way of dividing $R^n$ into two groups so
that $A$ and $B$ are contained in different groups ?

In mathematical terminologies, is there a way of obtaining a function $f$ : $%
R^n\longrightarrow R$ such that

\hspace{1.0in}$f(x)=\left\{ 
\begin{array}{c}
1,x\in A \\ 
-1,x\in B%
\end{array}
\right\} $ ?

The simplest function we may think of is a function of degree 1, i.e.,
linear function. Let's assume that $A$ and $B$ are separable by such a
function, which is of form $w\cdot x+b=0,$ where $x$ is a variable and $w$
and $b$ are parameters for the {\em hyperplane}. (Here $\cdot $ denotes the
standard inner product in Euclidean space.) Since there are infinitely many
hyperplanes separating $A$ and $B,$ for practical implementation, we need to
choose a specific one from those infinitely many candidates. Vapnik resolved
this problem elegantly by introducing the definition of the Optimal
hyperplane and connecting with nonlinear programming. It separate the sets $%
A $ and $B$ and the distance between the closest vectors of the two sets to
the plane is maximal.(For more details, see\cite{4}).

The key step of obtaining the optimal plane is to form a nonlinear
programming problem, by imposing appropriate restrictions, and apply the
Kuhn-Tucker's necessary conditions. This nonlinear programming problem makes
us choose a unique expression for a plane. For example, $0.5$ could be
expressed in infinitely many ways of fractions, i.e., $\frac{100}{200},\frac{%
28}{56},......\frac 48$. However, under the constraint that numerator and
denominator be relatively prime, then there is a unique fractional
expression, namely, $\frac 12.$ In the same principle, though there are
infinitely many planes for separation, in the respect of separating a data
set into exactly same two groups,(even for a single plane, it has infinitely
many expressions. $x+y=1$ and $2x+2y=2$ etc. represent the same line.), the
optimization problem deduced from an observation manages to get rid of this
ambiguity.

More precisely, it starts with the well-known distance formula from a point
to a plane and, for simplicity, let $n=2$.

Then the distance from a point $(x_{0,}y_0)$ to the plane $ax+by+c=0$ in the
plane is given by

\hspace{1.0in}$\frac{|ax_0+by_0+c|}{\sqrt{a^2+b^2}}.$

Observe that, if the condition $|ax_0+by_0+c|\geq 1$ is imposed, the
distance increases, as $\sqrt{a^2+b^2}$ decreases. Thus, if we have $(0,2)$
and $(0,-2),$ we have to solve the minimization problem under two
restriction. The line $\frac 12y=0$ will be the optimal hyperplane.
Intuitively, the optimal line should satisfy the maximal distance from both
points, which is the line passing through the middle point, the origin.

In general, let $(x_1,y_1),(x_{2,}y_{2,})......(x_l,y_l)$ be a set of
labelled vectors, where each $y_i$ denotes where $x_i$ belongs and takes
either $+1$ or $-1$. The formulation from the observation can be stated as
follows:

\hspace{1.0in}$minimize$ $f(w)=\frac 12||w||^2$

\hspace{1.0in}under the constraints, $y_i[(x_i\cdot w)+b]\geq 1,$ $%
i=1,2,....l.$:

One way of solving this optimization problem is to use the associated
Lagrangian whose definition is as follows:

\begin{definition}
Given the following optimization problem

$Minimize$ $f(x),$ $x=(x_1,x_2,....x_n),$

under the constraints $g_i(x)\geq 0,$ $i=1,2,...m,$

its associated Lagrangian is defined by

\hspace{1.0in}$L(x,\lambda )=f(x)-\sum_{i=1}^m\lambda _ig_i(x),$

where $\lambda =(\lambda _1,\lambda _2,.....\lambda _m)$, Lagrangian
multipliers.
\end{definition}

Kuhn and Tucker proved the minimization problem is equivalent to solving its
associated Lagrangian functional (See \cite{3}), i.e., finding a global
saddle point of its associated Lagrangian functional. In our case, the
associated Lagragian is given by

\hspace{1.0in}$L(w,b,\alpha )=\frac 12\Vert w\Vert ^2-\sum_{i=1}^l\alpha
_i[(x_i\cdot w+b)y_i-1]$

At the global saddle point, $L$ should be minimized with respect with to $w$
and $b,$ and maximized with respect to $\alpha _i\geq 0$. As a result, we
have familiar necessary conditions of first order derivatives, called
Kuhn-Tucker necessary conditions. Substitution those conditions in the
Lagrangian functional leads us to a quadratic programing:

\hspace{1.0in}$maximize$ $W(\alpha )=\frac 12\sum_{i,j=1}^ly_iy_j\alpha
_i\alpha _j(x_i\cdot x_j)-\sum_{i=1}^l\alpha _i$

\hspace{1.0in}under the constraints $\sum_{i=1}^l\alpha _iy_i=0$ and $\alpha
_i\geq 0,i=1,2...l.$

To construct a hyperplane of the optimal type in the case when the data set
is not separable linearly, we introduce non-negative {\em slack} variables $%
\varepsilon _i$'s to constraints to reduce the sum of ''distance of
separation'' errors.

\hspace{1.0in}$minimize$ $f(w)=\frac 12||w||^2+C\sum_{i=1}^l\varepsilon _i$

\hspace{1.0in}under the constraints, $y_i[(x_i\cdot w)+b]\geq 1-\varepsilon
_i,$ $i=1,2,....l.$:

Once again, the same arguments described above give the quadratic
programming,

\hspace{1.0in}$maximize$ $W(\alpha )=\frac 12\sum_{i,j=1}^ly_iy_j\alpha
_i\alpha _j(x_i\cdot x_j)-\sum_{i=1}^l\alpha _i$

\hspace{1.0in}under the constraints $\sum_{i=1}^l\alpha _iy_i=0$ and

\hspace{1.0in}$0\leq \alpha _i\leq C,$ $i=1,2...l.$ $C$ is a given constant.

The practical implementation of this quadratic optimization program was
discussed in some details.(See \cite{1})

\section{A representation of multiple SNP variations as a vector}

As we mentioned in {\bf section 1}, in this section, we propose a vector
representation of multiple SNP variations for the Support Vector Machine.
The basic concept is simple. As is often the case, comparison data with
other data is the best strategy and a reference is required. Height, weight,
blood alcohol concentration etc. are examples of association with numbers
that express quantities with respect to the standardize metric scale. Here
is a basic scheme we propose.

\begin{center}
{\bf Scheme}
\end{center}

Given each disease or trait, and a collection of SNP data which possibly
related to it,

1. Assume that there is no environmental factor.

2. SNP locations are assumed to be known for the disease or trait.

3. Assume there is a reference SNP\ data.

4. By giving scores based on {\em difference} from the reference data,
assign a vector to each SNP data. At each SNP location, take average of
those scores over the reference data set. The collection of {\em difference }%
scores, ordered numbers, may be considered as a vector in a Euclidean space,
whose dimension is the number of SNPs to be related to the disease or trait.

5. A training set is chosen for the disease or trait, in other words, SNP\
genotype data of normal and sick population.

6. By using {\bf Step} {\bf 4}, compute the SNP\ vectors of all training
data set. This would be a set of labelled vectors, $\{(x_i,y_i)\}$ as
described in {\bf section 2}

7. Use the Support Vector Machine to get a hyperplane dividing into two
groups, a control and a case group.

\begin{remark}
The reference data can be built by collecting SNP genotypes from the healthy
normal population. For example, choose 10000 ordinary and over 60 years old
people who have good health records. Depending on race, sex, region etc.,
the reference data should be distinguished into several different groups.
\end{remark}

\begin{remark}
At each SNP\ location, we might give a {\em difference} score uniformly. For
example,
\end{remark}

\hspace{1.0in}$diff(w/w,w/m)=$ $0.25,$ $diff(w/m,m/m)=0.75$ and

\hspace{1.0in}$diff(w/w,m/m)=1$

Here $w$ and $m$ represent {\em wild} and {\em mutation} genotypes
respectively.

\begin{remark}
The hyperplane obtained can be considered as a criterion, and, given a new
data set, it can be used for testing whether the person of the data is
susceptible to the disease or trait.
\end{remark}

\begin{remark}
Representation of an object as a vector might be critical for making use of
the Support Vector Machine. How to make domain knowledge contained in vector
representations is one of the major issues. For best performance, the
difference scores may be adjusted with training data set and, even at each
SNP location, such as giving weights on each SNP, normalization.We
implicitly used the fact those SNPs are related, i.e., linked each other. In
other words, they are dependent in the sense of statistical probability, but
independent in terms of linear algebra.(For this concept and an interesting
vector representation, see \cite{2})
\end{remark}

\begin{remark}
The idea of difference scoring could be applied to other data sets, in
particular, to haplotype data and to find out a linkage among SNP variations
etc.
\end{remark}

\begin{remark}
Once a group of SNP patterns are identified, it would be worth investigating
and computing contribution score of each of those SNP to the disease or
trait.
\end{remark}

\section{Inseparable Case}
For the inseparable case, the iterated use of support vector machine 
enables us to divide a collection of labelled vectors into several clusterring 
groups.

1. Set a threshold value. Say, 80 percent.

2. Use support vector machine to separate a collection of labelled vectors into two groups {\em A}, {\em B}.

3. Check if the groups contain more than 80 percent of either +1 or -1 labelled
vectors. Suppose {\em A} is not such one. Then use support vector machine to {\em A} again 
to separate into two subgroups. 

4. Repeat this procedure until each subgroup has a majority of more than 80 
percent. This will make each subgroup having a major set of labelled vectors.

5. For each subgroup, figure out a range, i.e., its center and radius . For example, averages and standard deviations of major labelled vectors may give information on the range.


\begin{thebibliography}{1}
\bibitem[1]{1} Joachims, Thorsten, Making large-Scale SVM Learning
Practical. Advances in Kernel Methods - Support Vector Learning, B. Sch\"{o}%
lkopf and C. Burges and A. Smola (ed.), MIT Press, 1999.

\bibitem[2]{2} Kim, MyungHo. A geometric model of information retrieval
systems, Complex Systems, Vol. 12, No. 1(2000), 93-102.

\bibitem[3]{3} Simmons, Donald, Nonlinear programming for operations
research, Prentice Hall, 1975

\bibitem[4]{4} Vapnik, Vladimir. The Nature of Statistical Learning,
Springer, New York, 1995
\end{thebibliography}
\end{document}